\documentstyle[preprint,aps]{revtex} 

\begin{document}
\preprint{}
\draft
\title{Mean Field Phase Diagram of $\rm SU(2)_L\times SU(2)_R$
 Lattice Higgs-Yukawa Model at Finite $\lambda$}
\author{Craig Pryor}
\address{Sierra Center for Physics\\
939 N. Van Ness Ave. \\
Fresno, CA 93728}
\date{}
\maketitle
\begin{abstract}
The phase diagram of an $\rm SU(2)_L\times SU(2)_R$
 lattice Higgs-Yukawa model with finite $\lambda$ is constructed using
mean field theory. The phase diagram bears a superficial resemblance
 to that for $\lambda=\infty$,
 however  as
 $\lambda$ is decreased the paramagnetic region shrinks in size.
For small $\lambda$ the phase transitions remain
second order, and no new first order transitions are seen.
\end{abstract}

\pacs{11.15.H, 11.15.E}
\newpage
\narrowtext

Recent experimental evidence of a top quark with mass $\approx 175$ GeV
\cite{topmass}
indicates
that its Yukawa coupling is of order $1$, possibly making
 nonperturbative effects  significant.
Lattice Higgs-Yukawa theories provide a way of studying
non-perturbative physics of theories containing interacting
scalars and fermions, although
the technical problems associated with chiral fermions restricts
us to vector theories.

Most of the work on Higgs-Yukawa theories
has dealt with the limit in which the Higgs field is radially frozen
($\lambda=\infty$). Such models have been studied
 both analytically \cite{stephanov,steph1st,O4mfa2}
and using Monte Carlo \cite{mclambinf} for various gauge groups.
Radially active (finite $\lambda$)
 Higgs models without fermions have also been thoroughly
examined \cite{mclambfin,anallambfin},
however relatively little work has been done on the problem
including fermions \cite{lambfinwferm}.
In this paper I estimate the phase diagram of  the
$SU(2)_L\times SU(2)_R$
Higgs-Yukawa theory
for the full range of $\lambda$ and the Yukawa coupling.
The calculations are performed using the mean field approximation
 (MFA)\cite{mfa}, with the fermions included in the manner
 used by Stephanov and Tsypin for the radially frozen
theory\cite{steph1st}.

The action for the model  in $d$ dimensions is \hbox{$S=S_F+S_H+V_H$},
\begin{eqnarray}\nonumber
S_F={1\over 2} \sum_{x,\mu}
   (\bar \psi_x       \gamma_\mu   \psi_{x+\mu}
    -\bar \psi_{x+\mu} \gamma_\mu \psi_{x    } )\nonumber\\
+y\sum_x  \bar  \psi_x (P_R^{~} \rho_x \Phi_x^{~}+P_L^{~} \rho_x
      \Phi_x^{\dagger})  \psi_x\nonumber\\
=\sum_{xy}\bar \psi_x [K_{xy}+y  (P_R^{~} \rho_x \Phi_x^{~}+P_L^{~}
   \rho_x \Phi_x^{\dagger})\delta_{xy}] \psi_y\nonumber\\
=\sum_{xy}\bar \psi_x M_{xy} \psi_y\label{SF}\\
   S_H=-\kappa \sum_{x,\mu}{1\over 2}
   {\rm tr}(\rho_x^{~}\Phi_x^\dagger  \rho_{x+\mu}\Phi_{x+\mu}+
    \rho_{x+\mu}^{~}\Phi_{x+\mu}^\dagger  \rho_x \Phi_{x})\\
V_H= \sum_{x}V(\rho_x)= \sum_{x} \lambda(\rho_x^2 -1)^2+\rho_x^{~}
.\label{SH}
\end{eqnarray}
The fermion
field $\psi_x$ is an SU(2) doublet,
$P_{R,L}={1\over 2}(1\pm \gamma_{d+1}^{~})$,
and $\kappa$ is the scalar hopping parameter.
The scalar field has been separated into its magnitude and angular
degree of freedom, $\rho_x$ and $\Phi_x$ respectively,
where $\Phi_x^{~}$ is a $2\times 2$ SU(2) matrix.
We will also use the
 $O(4)$ field $\phi^k_x$ satisfying
$\phi^k_x \phi^k_x =1$.
The two notations are related by
 $\Phi_x^{~}=\phi^k_xT^k_{~}$ where $T^k_{~}=({\openone},i\vec \sigma)$.
Integrating out the fermions gives the effective action for the
scalar field
\begin{equation}
S_{eff}=S_H+V_H-N_F^{~}\ln \det M, \label{Seff}
\end{equation}
where $N_F^{~}$ has been introduced as the number of fermion species,
each
having the same action $S_F$.
The partition function for the effective scalar theory is
\begin{equation}Z=\int\prod_x\rho^3_xd\rho_x\int \prod_xD\phi_x
\exp(-S_{eff}) \label{Z}
\end{equation}
where $D\phi_x$ is the $O(4)$ invariant group measure for $\phi^k_x$.

The variational form of the MFA is employed
 by introducing a  parameter $H^k$, then
adding and subtracting the trial
action $\rho_x \phi_x^k H^k$ to obtain
\begin{equation}\exp(-F)=\int\prod_x\rho^3_xd\rho_x\int \prod_xD\phi_x
\exp(-S_{eff}+\sum_x\rho_x \phi_x^k H^k-\sum_x\rho_x \phi_x^k H^k).
\label{F}
\end{equation}
This gives the variational limit on the free energy $F$,
\begin{eqnarray}
  F\le F_{var}=\langle S_H-N_F \ln \det M\rangle_H+
  \langle \sum_x H^k\phi^k_x \rho_x \rangle_H
   -\ln Z_H\label{Fvar},\\
\langle A \rangle_H=Z_H^{-1}
  \int \prod_x d\rho_x \rho_x^3\exp (-V(\rho_x))
  \int \prod_x D\phi_x
  \exp(\sum_x H^k\phi_x^k \rho_x)
  A\label{avg},\\
Z_H=\int \prod_x d\rho_x \rho_x^3\exp (-V(\rho_x))
  \int \prod_x D\phi_x \exp(\sum_x H^k\phi_x^k \rho_x).\label{ZH}
\end{eqnarray}
$F_{var}$ is then minimized  with respect to $H^k$.
The integration over $\rho_x$ is necessary since
for finite $\lambda$
the magnitude of the scalar is no longer constrained to be $1$.
At small $\lambda$ the
shallow scalar potential causes fluctuations about $\rho_x=1$ which
alter the results considerably.

For small Yukawa coupling the fermionic determinant  is calculated by
expanding in $y$
\begin{eqnarray}
\langle \ln \det M \rangle_{H}=
\ln \det K_{xy}-\sum_{n=2,4,...}
{1\over n}y^n\sum_{x_1...x_n}K_{x_1x_2}^{-1}...K_{x_nx_1}^{-1}
\langle {\rm tr}\Phi_{x_1}^{~}
\Phi_{x_2}^\dagger...\Phi_{x_{n-1}}^{~}\Phi_{x_n}^\dagger
\rangle_H\nonumber\\
\approx\ln \det K_{xy}-\sum_{n=2,4,...}
{1\over n}\bigg({-2y\over d}\bigg)^n\sum_{x_1...x_n}
K_{x_1x_2}^{~}...K_{x_nx_1}^{~}
\langle {\rm tr}\Phi_{x_1}^{~}
\Phi_{x_2}^\dagger...\Phi_{x_{n-1}}^{~}\Phi_{x_n}^\dagger
\rangle_H.\label{expdet}
\end{eqnarray}
The second line uses the approximation\cite{steph1st}
\begin{equation}
K^{-1}_{xy}={-2\over d}K^{~}_{xy} + O(1/{d^2}), \label{propappr}
\end{equation}
which is justified
since the MFA is itself only accurate to order $1/d$.
The determinant is then represented by a sum
of closed hopping diagrams connecting sites
associated with alternating $\Phi$'s and $\Phi^\dagger$'s.
Previous work has approximated the fermionic determinant
by including an infinite subset of hopping diagrams
 \cite{steph1st}, or expanding Eq.
(\ref{expdet}) to some finite order in $y$ \cite{O4mfa2}. We will adopt
the latter approach.

The
$\lambda =\infty$ case provides a hint in
choosing how far to carry out the expansion of the fermionic determinant.
Since the radially frozen theory in the MFA agrees quite well with Monte
Carlo results
when the free energy is expanded to order $y^4$,
 we will evaluate Eq. \ref{expdet} to order $y^4$.
It is a tedious though straightforward exercise to enumerate all such
hopping diagrams in $d$ dimensions to obtain
\begin{eqnarray}
\langle \ln \det M \rangle_{H}
\approx \ln \det K +{2^{d/2}\over d}y^2  {\rm tr} \langle \rho_x \Phi_x
\rangle_H
\langle \rho_x\Phi_x \rangle_H- \nonumber \\
{2^{d/2}\over 2 d^3}y^4
\bigg[ {\rm tr}\langle \rho_x\Phi_x^{~}\rho_y\Phi_y^\dagger\rho_x\Phi_x^{~}
\rho_y\Phi_y^\dagger
\rangle_H   \nonumber\\
+2(2d-1){\rm tr}\langle\rho_x\Phi_x^{~}\rho_x\Phi_x^{~}\rangle_H
 \langle\rho_x\Phi_x^{~}\rangle_H  \langle\rho_x\Phi_x^{~}\rangle_H\nonumber\\
+2(d-1){\rm tr} \langle \rho_x\Phi_x^{~}\rangle_H^4
\bigg] \label{answer}
\end{eqnarray}
where the traces are over $SU(2)$ indices,
and the indices $x$ and $y$ simply indicate
which $\Phi$'s are distinct group elements.

The various quantities of the form $\langle A \rangle_H$
 in Eq. \ref{answer}
include group integrals, and integrals over $\rho$.
The group integrals are calculated to second order in
$H$ because that is all that is required to find second
order critical lines.
We first compute
the  $O(4)$ group integrals by taking
derivatives with respect to $H$ of
\begin{eqnarray}
\int D\phi_x \exp( H^n\phi_x^n \rho_x)=
  2\pi^2[I_0(H\rho_x)-I_2(H\rho_x)] \nonumber\\
=2\pi^2 (1+{1\over 8}\rho_x^2 H^2+{1\over 192}\rho_x^4 H^4)+O(H^6)
\label{genfxn}
\end{eqnarray}
where $H=\sqrt {H^kH^k}$ and $I_n(x)$ is the $n$th order modified
Bessel function. We find
\begin{eqnarray}
\int D\phi_x \exp(\rho_x H^n\phi_x^n)
 \phi^i_{x} = {\pi^2\over 2} \rho_x  H^i+O(H^3),\\     \label{gpint1}
\int D\phi_x \exp(\rho_x H^n\phi_x^n)
 \phi^i_{x} \phi^j_x
= {1\over 12}\pi^2 \rho_x^2 H^i H^j +\delta^{ij}({1\over 2}+{1\over 24}
\rho_x^2 H^2)\pi^2+O(H^4)  \label{gpint2}.
\end{eqnarray}
The corresponding $SU(2)$ group
integrals are
\begin{eqnarray}
{\rm tr}\langle \Phi_x \Phi_x \rangle_H=-2\pi^2+O(H^4),\label{gpint3}\\
\langle \Phi_x \rangle_H= \openone {\pi^2\over 2}H\rho_x+O(H^3).
\label{gpint4}
\end{eqnarray}
In Eq. \ref{gpint4} the gauge has been fixed to $H^i=H\delta^{i0}$.

In addition, the slightly more complicated group integral involving
four $\Phi$'s is needed.
This is calculated  as
\begin{eqnarray}
{\rm tr}\int D\phi_x D\phi_y \exp(\rho_x H^n \phi^n_x+\rho_y H^n \phi^n_y)
\Phi_x\Phi^\dagger_y\Phi_x\Phi^\dagger_y\nonumber\\=
(\delta^{ij}\delta^{kl}+\delta^{il}\delta^{jk}-\delta^{ik}\delta^{jl})
\int D\phi_x  \exp(\rho_x H^n \phi^n_x)\phi^i_x\phi^k_x
\int D\phi_y  \exp(\rho_y H^n \phi^n_y)\phi^j_y\phi^l_y\nonumber\\
=-\pi^4 (4+{1\over 2}H^2\rho_x^2+{1\over 2}H^2\rho_y^2)+O(H^4).
\end{eqnarray}
The second line
relies on the fact that the trace
 must be separately
symmetric in both $i,k$ and in $j,l$.
The  sum over $O(4)$ indices  is computed using Eq. \ref{gpint1} and
by again choosing the $H^i=H\delta^{i0}$ gauge.

For computing the integrals over $\rho$
it is useful to define
\begin{eqnarray}
    P_n=\int_0^\infty d\rho \rho^n \exp(-V(\rho))\nonumber\\
={1\over2} (2\lambda)^{-{n+1\over 4}}
\Gamma({n+1\over 2})
D_{-{n+1\over 2}}\left({(1-2\lambda )\over\sqrt {2\lambda}}\right)
\exp\left( {1\over 8\lambda}- {1\over 2}-{\lambda\over 2}   \right)
\end{eqnarray}
where     $D_{-{n+1\over 2}}$ is the parabolic
cylinder function.
The single site partition function is then given by
\begin{eqnarray}
   Z_1=2\pi^2\int_0^\infty d\rho_x \rho_x^3
   \exp(-V(\rho_x))[I_0(\rho_x H)-I_2(\rho_x H)]\nonumber\\
= 2\pi^2 \int_0^\infty d\rho_x
    \rho_x^3 \exp(-V(\rho_x))[1+{1\over8}H^2\rho_x^2]+O(H^4)\nonumber\\
= 2\pi^2(P_3+{1\over 8}H^2 P_5)+O(H^4).
\end{eqnarray}
Similarly, the integrals making up Eq. \ref{Fvar} are given by
\begin{eqnarray}
    \langle \rho_x \Phi \rangle_H=\openone {1\over 2}\bigg[{ P_5\over 2P_3}H
    +O(H^3)\bigg]\label{phi},\\
   {\rm tr} \langle \rho_x \Phi \rho_x \Phi \rangle_H=
{-P_5\over P_3}+{P_5^2\over 8P_3^2}H^2+O(H^4)\label{phi2},\\
   {\rm tr} \langle \rho_x\Phi_x^{~}
              \rho_y\Phi_y^\dagger
              \rho_x\Phi_x^{~}
              \rho_y\Phi_y^\dagger
   \rangle_H=
    \bigg({P_5\over P_3}\bigg)^2+H^2{P_3P_5P_7-3P_5^3\over 12P_3^3}+O(H^4).
\label{phi4}
\end{eqnarray}
Substituting Eq.'s \ref{phi}-\ref{phi4} into Eq.'s
 \ref{answer} and \ref{Fvar} gives $F_{var}$.

There are four phases: 1) the ferromagnetic (FM)
 phase with $\langle \phi^i_x \rangle \ne 0$, 2) the paramagnetic
(PM)
phase with $\langle \phi^i_x \rangle =0$, 3) the
antiferromagnetic (AFM) phase with
$\langle\xi_x\phi^i_x \rangle\ne 0$ (where
$ \xi_x =(-1)^{x_1+x_2+..+x_d}$), and 4) the ferrimagnetic (FI) phase
in which  $\langle \phi^i_x \rangle \ne 0$ and
$\langle\xi_x\phi^i_x \rangle\ne 0$.
In the MFA these phases are indicated by the value of $H^k$ which
minimizes
the variational free energy. In the FM phase  $H^k\ne0$, while
for the PM phase $H^k=0$.
To see the AFM phase the action must be transformed so as to make the
staggered magnetization $\langle\xi_x\phi^i_x \rangle$ accessible.
Since the  action is invariant under
$\phi_x \rightarrow \xi_x \phi_x$,
$\psi_x \rightarrow \exp (i\pi 2 \xi_x /2)\psi_x $,
$y \rightarrow iy$,
$\kappa \rightarrow -\kappa$,
the action obtained by this transformation will have AFM order if
$H^k\ne0$.
Therefore, the MFA variational action for the AFM is given by Eq. \ref{answer}
with $y \rightarrow iy$, $\kappa \rightarrow -\kappa$.
The FI phase is indicated by the existence of FM and AFM order at the same
point.

The strong coupling regime is reached by
 expanding the fermionic determinant in $1/y$ rather than $y$.
In this expansion $K_{xy}$ appears rather than  $K_{xy}^{-1}$ so the
$1/d$ approximation of the propagator is not needed.
 Aside from this difference, the free energy is
computed in the same manner
 as the weak coupling version. Therefore the strong coupling
variational free energy is obtained
 making the replacement $y\rightarrow d/2y$ in Eq. \ref{answer}.

The second order phase transitions are found by setting
${\partial^2\over \partial H^2}F_{var}|_{H=0}=0$, which is
why it was sufficient to compute $F_{var}$ to $O(H^2)$. For weak
coupling this gives
\begin{eqnarray}
{\rm FM-PM:}{~~}\kappa_c={P_3\over P_5d}-{2^{d/2}N_F \over d^2}y^2+
{2^{d/2}N_F P_5\over P_3d^3}\bigg({1\over 2 d}-1\bigg)y^4,\\
{\rm AFM-PM:}{~~}\kappa_c=-{P_3\over P_5d}-{2^{d/2}N_F \over d^2}y^2-
{2^{d/2}N_F P_5\over  P_3d^3}\bigg({1\over 2 d}-1\bigg)y^4.
\end{eqnarray}
The figures illustrate the above results.
 Fig. 1 shows $\kappa_c$
as a function of $\lambda$ at $y=0$ for the FM-PM transition.
(At $y=0$, $\kappa_c$ for the AFM-PM transition is simply the
negative of $\kappa_c$ for the FM-PM transition.)
In agreement with previous results  \cite{mclambfin,purehiggs},
$\kappa_c\rightarrow 1/8$  as $\lambda \rightarrow 0$,
and
$\kappa_c\rightarrow 1/4$  as $\lambda \rightarrow \infty$.
Fig. 2 shows the complete phase diagram for $\lambda=1$, which
appears qualitatively similar to the $\lambda=\infty$ case. The
one difference from the $\lambda=\infty$ case is that the PM region
has shrunk.
Fig. 3 shows the
$\kappa_c$'s as a function of $y$ for several values of $\lambda$
between $0.01$ and $10$.
As $\lambda$ is decreased, the phase diagram remains qualitatively the same
as the $\lambda=\infty$ case, although
the PM region shrinks in both the $y$ and $\kappa$ directions.
All phase transitions remain second order, and there is no evidence
of new phases. Also,
$\kappa_c$ saturates for $\lambda=0.01,10$, with little change in
$\kappa_c$ as $\lambda \rightarrow 0,\infty$.

In addition to the MFA errors of order $1/d^2$, the errors in the
expansion of the fermionic determinant are
of order $y^6$. For $y\approx 1$ the results should
break down, although comparison of the $\lambda=\infty$ results with
those from Monte Carlo show good agreement through the intermediate
coupling region.

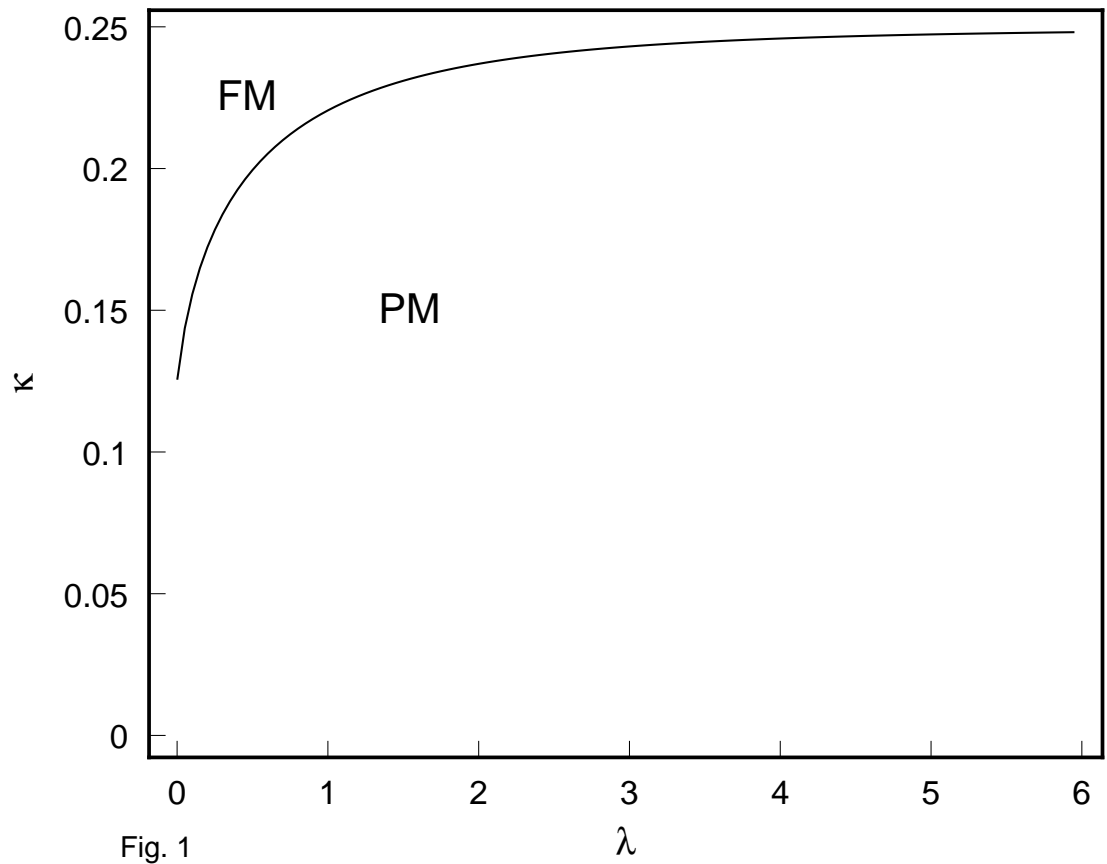
\begin{figure}
\caption{$\kappa_c$ for the FM-PM transition at $y=0$.
}
\label{fig1}
\end{figure}

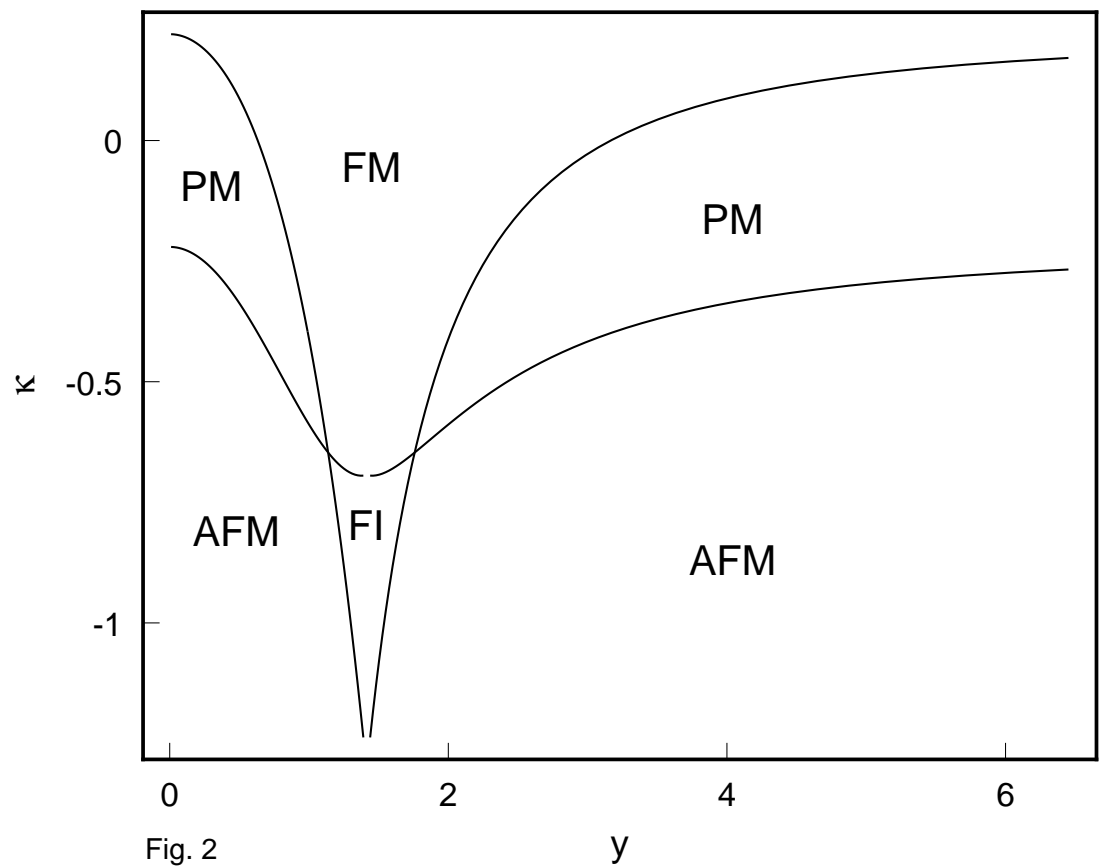
\begin{figure}
\caption{The phase diagram for $\lambda=1$. All phase
transitions are second order.
}
\label{fig2}
\end{figure}

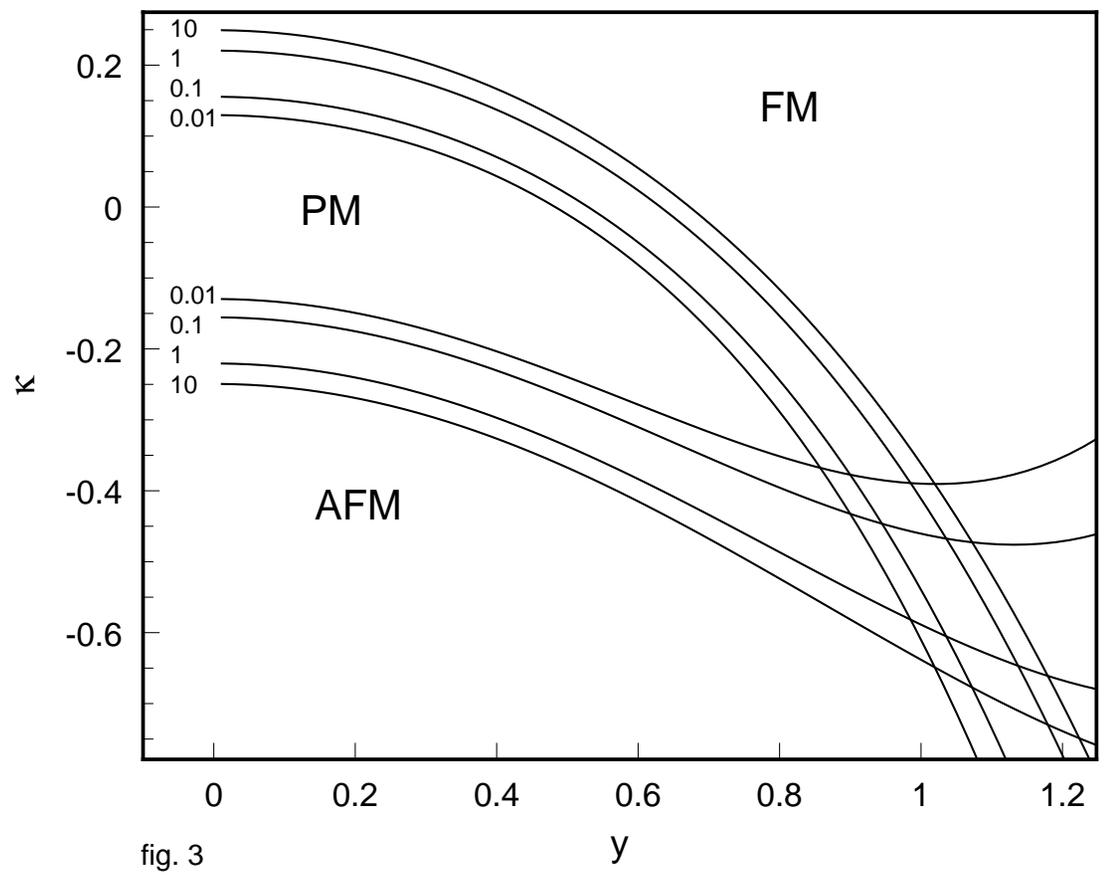
\begin{figure}
\caption{Second order $\kappa_c$'s for various values of $\lambda$
as indicated by the labels.
The $\lambda =0.01, 10$ lines are very nearly the same as those for
$\lambda=0,\infty$ respectively.
}
\label{fig3}
\end{figure}


\begin{references}

\bibitem{topmass} CDF Collaboration, F. Abe {\it et. al.}, Phys.\ Rev.\ D
  {\bf 50}, 2966 (1994);
  Phys.\ Rev.\ Lett.\ {\bf 73}, 225 (1994).
\bibitem{stephanov}
  M. A. Stephanov, M. M. Tsypin, Phys.\ Lett.\  {\bf 261B}, 109 (1991);
  ibid. {\bf 242B}, 432 (1990);
  ibid. {\bf 236B}, 344 (1990).
\bibitem{steph1st}
  M. A. Stephanov, M. M. Tsypin, Zh.\ Eksp.\ Teor.\ Fiz.\ {\bf 97}, 409 (1990).
\bibitem{O4mfa2}
  T. Ebihara, K. Kondo, Prog.\ Theor.\ Phys. {\bf 87}, 1019 (1992).
\bibitem{mclambinf}
  W. Bock {\it et. al.}, Nucl.\ Phys.\ {\bf 344}, 207 (1990);
  W. Bock, A. K. De, Phys.\ Lett.\ {\bf 245B}, 207 (1990).
\bibitem{mclambfin}
  H. Kuhnelt, C.B. Lang, G. Vones,  Nucl.\ Phys.\ B {\bf 230}, 16 (1984);
  I. Montvay, Phys.\ Lett.\ {\bf 150B}, 441 (1985);
  J. Jersak, et. al., Phys.\ Rev.\ D {\bf 32}, 2761 (1985).
\bibitem{anallambfin}
  R. E. Shrock,  Phys.\ Lett.\ {\bf 180B}, 268 (1986);
  P. H. Damgaard, U. M. Heller,  Phys.\ Lett.\ {\bf 164B}, 121 (1985);
  Y. Sugiyama, T. Yokota,  Phys.\ Lett.\ {\bf 168B}, 386 (1986);
  Y. Sugiyama, T. Yokota,   Prog.\ Theor.\ Phys. {\bf 76}, 667 (1986);
  T. Munehisa, Y. Munehisa, Nucl.\ Phys.\ B {\bf 215}, 508 (1983);
  Y. Sugiyama, K. Kanaya,   Prog.\ Theor.\ Phys. {\bf 73}, 176 (1985).
\bibitem{lambfinwferm}
  A. Hasenfratz, K. Jansen, Y. Shen,  Nucl.\ Phys.\ B {\bf 394}, 527 (1993).
\bibitem{mfa}
  J. M. Drouffe, J. B. Zuber, Phys.\ Rep.\ {\bf 102}, 1 (1983).
  and references contained therein.
\bibitem{purehiggs}
   Hasenfratz et. al., Nucl.\ Phys.\ B {\bf 365}, 79 (1991);
   P. R. Gerber, M. E. Fisher, Phys.\ Rev.\ B {\bf 10}, 4697 (1974);
   K. Jansen, et. al., Nucl.\ Phys.\ B {\bf 265}, 129 (1986).

\end{references}
\end{document}